\documentclass[aps,prl,twocolumn,showpacs,superscriptaddress,groupedaddress]{revtex4}
\usepackage{epsfig}
\usepackage{graphicx}
\usepackage{bm}
\usepackage{amsmath}

\usepackage[colorlinks=true, pdfstartview=FitV, linkcolor=red, citecolor=blue, urlcolor=blue]{hyperref}

\usepackage{color}

\begin{document}

\title{Elliptic and triangular flow in p+Pb and peripheral Pb+Pb collisions from\\parton scatterings}

\author{Adam Bzdak}
\email{abzdak@quark.phy.bnl.gov}
\affiliation{RIKEN BNL Research Center, Brookhaven National Laboratory, Upton, NY 11973,
USA}
\affiliation{AGH University of Science and Technology, 30-059 Krakow,  Poland}

\author{Guo-Liang Ma}
\email{glma@sinap.ac.cn}
\affiliation{Shanghai Institute of Applied Physics, Chinese Academy of Sciences, Shanghai 201800, China}

%\date{\today }

\begin{abstract}
Using a multiphase transport model (AMPT) we calculate the elliptic, $v_2$, and triangular, $v_3$, Fourier coefficients of the two-particle azimuthal correlation function in proton-nucleus (p+Pb) and peripheral nucleus-nucleus (Pb+Pb) collisions. Our results for $v_3$ are in a good agreement with the CMS data collected at the Large Hadron Collider. The $v_2$ coefficient is very well described in p+Pb collisions and is underestimated for higher transverse momenta in Pb+Pb interactions. The characteristic mass ordering of $v_2$ in p+Pb is reproduced whereas for $v_3$ this effect is not observed. We further predict the pseudorapidity dependence of $v_2$ and $v_3$ in p+Pb and observe that both are increasing when going from a proton side to a Pb-nucleus side. Predictions for the higher order Fourier coefficients, $v_4$ and $v_5$, in p+Pb are also presented.
\end{abstract}

\maketitle

\begin{figure*}[!ht]
\begin{center}
\includegraphics[scale=0.7]{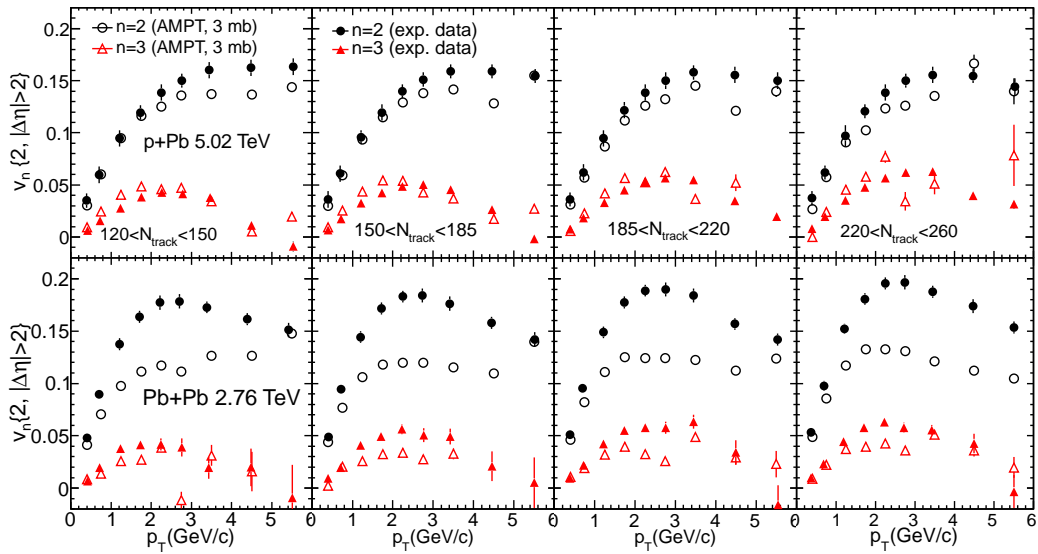}
\end{center}
\par
\vspace{-5mm}
\caption{The transverse momentum dependence of the elliptic, $v_2$, and triangular, $v_3$, flow coefficients in p+Pb (upper panel) and Pb+Pb collisions (lower panel) as obtained in the AMPT model (open symbols) with the string melting mechanism. Different centrality classes are defined by the number of produced charged particles, $N_{\rm track}$, measured in $|\eta|<2.4$ and $p_{\rm{T}}>0.4$ GeV/$c$. The CMS data are denoted by the full points.}
\label{fig:v2_v3_pt}
\end{figure*}

\begin{figure*}[!ht]
\begin{center}
\includegraphics[scale=0.6]{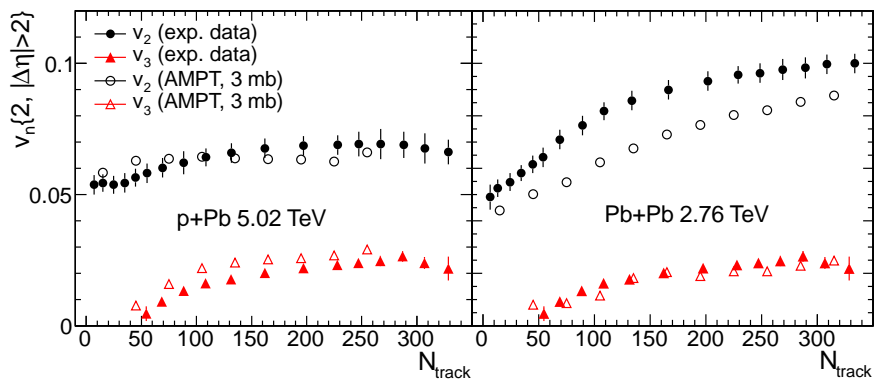}
\end{center}
\par
\vspace{-5mm}
\caption{The CMS data (full points) vs the AMPT model (open symbols) with the string melting mechanism for the integrated elliptic, $v_2$, and triangular, $v_3$, flow coefficients in p+Pb (left) and Pb+Pb (right) collisions as a function of the number of produced charged particles, $N_{\rm track}$, measured in $|\eta|<2.4$ and $p_{\rm{T}}>0.4$ GeV/$c$.}
\label{fig:v2_v3}
\end{figure*}

Recently we argued \cite{Ma:2014pva} that the incoherent scattering of partons, as present in a multiphase transport model (AMPT) \cite{Lin:2004en}, with a modest elastic parton-parton cross-section $\sigma = 1.5-3$ mb, allows to understand qualitatively and quantitatively the long-range two-particle azimuthal correlation functions in proton-lead (p+Pb) and high-multiplicity proton-proton (p+p) collisions. Such correlations were recently observed by the CMS \cite{Khachatryan:2010gv,CMS:2012qk,Chatrchyan:2013nka}, ALICE \cite{Abelev:2012ola,ABELEV:2013wsa} and ATLAS \cite{Aad:2012gla,Aad:2013fja} Collaborations at the Large Hadron Collider (LHC), and by the PHENIX Collaboration in deuteron-gold collisions at the Relativistic Heavy Ion Collider (RHIC) \cite{Adare:2013piz}.

Interestingly all features of the two-particle azimuthal correlation function observed in p+Pb collisions are very similar to those observed in A+A interactions, where such correlations are commonly attributed to the hydrodynamic expansion of the produced fireball. This naturally suggests that collective physics is present in p+A collisions \cite{Bozek:2011if,Shuryak:2013ke,Bzdak:2013zma,Bozek:2013uha,Qin:2013bha,Werner:2013tya,Kozlov:2014fqa,Ma:2014pva}. Particularly strong evidence in favour of hydrodynamics (or any other approach where the initial coordinate space anisotropy is transformed into the final momentum anisotropy) in p+A and peripheral A+A collisions is an approximate equality of multi-particle elliptic flow cumulants, $v_2\{4\} \approx v_2\{6\} \approx v_2\{8\}$, as predicted in Ref. \cite{Bzdak:2013rya}, see also \cite{Yan:2013laa,Bzdak:2013raa}, and confirmed recently by the CMS Collaboration \cite{CMS:v4-v8:qm14:talk}. Other strong evidence is the characteristic mass ordering of the elliptic flow, $v_2$, observed by the ALICE Collaboration in Ref. \cite{ABELEV:2013wsa}, and successfully reproduced by hydrodynamic calculations \cite{Bozek:2013ska,Werner:2013ipa}.

The experimental data for the two-particle azimuthal correlation function can be also fitted within the color glass condensate framework \cite{Gelis:2010nm}, where the interesting part of the two-particle correlation function comes from the emission of two gluons in the so-called glasma diagram \cite{Dusling:2013oia}. For a detailed discussion of this approach we refer the reader to Refs. \cite{Dusling:2013oia,Kovchegov:2012nd,Kovner:2012jm}.

It is important to clarify whether the signal in p+A collisions comes form the initial or final (or both) state effects. To this end several interesting observations were recently published \cite{Bzdak:2013lva,Bzdak:2013zla,Bozek:2013sda,Basar:2013hea,Konchakovski:2014wqa,Sickles:2013yna,Noronha:2014vva, Floerchinger:2014fta} which could help to distinguish between competing models of p+A interactions. A simple conformal scaling argument, presented in Ref. \cite{Basar:2013hea}, indicates a presence of a collective response to the geometry in p+Pb and Pb+Pb collisions.
 
In this Letter we focus on the detailed discussion of the elliptic and triangular coefficients of the two-particle azimuthal correlation function in p+Pb and peripheral Pb+Pb collisions. Qualitatively we reproduce all trends observed in the data. In particular, we find that $v_2$ in p+Pb is in a good agreement with the CMS data for a broad range of $N_{\rm track}$ and $p_{\rm{T}}$. In peripheral Pb+Pb collisions $v_2$ is underestimated for higher $p_{\rm{T}}$ and the integrated $v_2$ is $20\%$ below the data (the model is not expected to work with better accuracy) however, the $N_{\rm track}$ functional dependence is well reproduced. As far as the $v_3$ coefficient \cite{Alver:2010gr} is concerned we obtain a good description of the data in both p+Pb and Pb+Pb collisions. We observe that the integrated $v_3$ is very similar in both systems for a broad range of $N_{\rm track}$. We further confirm the mass ordering of $v_2$ which is a characteristic feature of collective dynamics. Finally we predict the dependence of the two-particle correlation function on the pseudorapidity sum, $\eta_1 + \eta_2$, at a given pseudorapidity separation, $\eta_1 - \eta_2$, between two particles. We observe that $v_2$ and $v_3$ increase with the rapidity sum (that is when going towards a Pb fragmentation region), which is thought as a helpful probe to distinguish between various models of p+Pb collisions. We also predict the higher-order Fourier coefficients, $v_4$ and $v_5$, in p+Pb collisions and find them roughly a factor of $2$ smaller than the $v_3$ coefficient.

Similar to our previous work we use the AMPT model with the string melting mechanism. In this model all initial minijets and soft strings are converted into quarks and antiquarks which undergo elastic scatterings (in contrast to the default model, where only partons from minijets interact) with a partonic cross-section which is controlled by the strong coupling constant and the Debye screening mass. Subsequently a simple coalescence model is employed to form hadrons which further undergo hadronic scatterings. The detailed description of the AMPT model can be found in Ref. \cite{Lin:2004en}. The AMPT model provides a consistent framework to understand many phenomena in p+p, p+A and A+A collisions. In particular, different orders of harmonic coefficients have been well reproduced in Au+Au collisions at the top RHIC energy~\cite{Ma:2014xfa} and Pb+Pb collisions at the LHC energy~\cite{Xu:2011jm}, which indicates that in A+A interactions, the initial spacial asymmetry is transformed into the final momentum anisotropy via the incoherent parton scatterings \cite{Ma:2010dv}. 

In our previous study, the long-range two-particle azimuthal correlations have been observed in p+p and p+Pb collisions at the LHC energies with a modest parton-parton cross-section of $\sigma=1.5 - 3$ mb~\cite{Ma:2014pva}. Therefore, it is important to check if the flow coefficients $v_{n}$ extracted from the long-range two-particle azimuthal correlation function are comparable with the data. In this work we simulate p+Pb collisions at $\sqrt{s}$ = 5.02 TeV and peripheral Pb+Pb collisions ($50-100\%$) at $\sqrt{s}$ = 2.76 TeV with the parton-parton cross-section of $3$ mb, being consistent with our previous study. 

In Fig. \ref{fig:v2_v3_pt} we present the elliptic and triangular Fourier coefficients from the long-range two-particle azimuthal correlation functions, i.e. $v_{n}\{2,|\Delta \eta|>2\}$, as a function of the transverse momentum, $p_{\rm{T}}$, in p+Pb (upper panel) and Pb+Pb collisions (lower panel) at $\sqrt{s} = 5.02$ TeV  and $2.76$ TeV, respectively. In our analysis we exactly follow the CMS procedure as described in Ref. \cite{Chatrchyan:2013nka}. The description of the p+Pb data is very good for both $v_2$ and $v_3$ in the whole available transverse momentum range and for various centrality classes defined by the number of produced charged particles, $N_{\rm track}$, measured in $|\eta|<2.4$ and $p_{\rm{T}}>0.4$ GeV/$c$. This is a nontrivial result suggesting that the AMPT model captures the main features of p+A physics. In Pb+Pb collisions $v_3$ is consistent with the data, within the error bars, and surprisingly $v_2$ is underestimated for $p_{\rm{T}}>1$ GeV/$c$ \cite{footnote}. It is interesting to notice that $v_2(p_{\rm{T}})$ in Pb+Pb has a characteristic maximum around $p_{\rm{T}} = 2.5$ GeV/$c$ which is not present in p+Pb data. On the contrary $v_3(p_{\rm{T}})$ is very similar in both systems and is well described by the AMPT model.

In Fig. \ref{fig:v2_v3} we present the integrated ($0.3<p_{\rm{T}}<3$ GeV/$c$) $v_2$ and $v_3$ for both p+Pb and Pb+Pb collisions. Again, $v_2$ and $v_3$ are very well described in p+Pb collisions for all available $N_{\rm track}$. Unfortunately, at present we cannot go to the highest values of $N_{\rm track}>300$ to check whether $v_3$ starts decreasing as suggested by the data. In Pb+Pb collisions the integrated $v_3$ is consistent with the data for all $N_{\rm track}$ and the $v_2$ coefficient is underestimated by roughly $20\%$. It is worth noticing that within the AMPT approach the integrated $v_3$ in p+Pb and Pb+Pb interactions is roughly the same.

\begin{figure}[t]
\begin{center}
\includegraphics[scale=0.4]{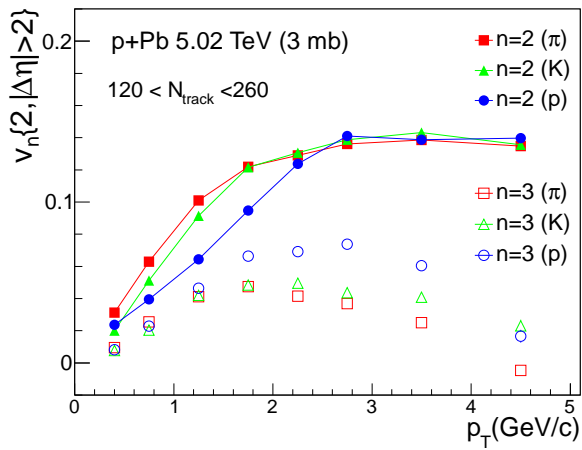}
\end{center}
\par
\vspace{-5mm}
\caption{The elliptic and triangular flow coefficients in p+Pb ($120 < N_{\rm track} < 260$) as a function of the transverse momentum for pions, kaons and protons as obtained in the AMPT model with the string melting mechanizm.}
\label{fig:v2_v3_pi_K_p}
\end{figure}

It is interesting to calculate $v_2(p_{\rm{T}})$ and $v_3(p_{\rm{T}})$ separately for pions, kaons and protons. The recently observed mass ordering of $v_2$ in p+Pb collisions serves as a crucial test of the initial vs the final state effects. In hydrodynamics we naturally obtain the mass ordering \cite{Bozek:2013ska,Werner:2013ipa} which is not obvious in the initial state scenarios.  
We checked that the mass ordering of $v_2$ is present in the AMPT model, as presented in Fig. \ref{fig:v2_v3_pi_K_p}. Interestingly we do not observe the mass ordering for $v_3$, being consistent with the calculations of Ref. \cite{Bozek:2013ska}. 

\begin{figure}[!h]
\begin{center}
\includegraphics[scale=0.48]{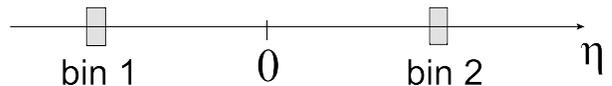}
\end{center}
\par
\vspace{-5mm}
\caption{Two narrow bins in pseudorapidity with $\Sigma \eta = \eta_1 + \eta_2 \sim 0$. We shift both bins simultaneously to study the dependence of $C(\Delta\phi)$ on $\Sigma \eta$ at a given $\Delta \eta = \eta_2 - \eta_1 $.}
\label{fig:2_bins}
\end{figure}

We further present our predictions for the pseudorapidity dependence of the two-particle azimuthal correlation function in p+Pb collisions. In our calculations we take two narrow pseudorapidity bins with a given pseudorapidity separation $\Delta \eta = \eta_2 - \eta_1$. Next we shift both bins simultaneously across the pseudorapidity axis to study the azimuthal correlation function for various values of the pseudorapidity sum, $\Sigma \eta = \eta_1+\eta_2$, at a given $\Delta \eta$. Schematically this situation is presented in Fig. \ref{fig:2_bins}. We calculate the two-particle azimuthal correlation function, $C(\Delta \phi)$, defined as
\begin{equation}
C(\Delta \phi) \equiv 
\frac{Y_{\rm Same}(\Delta\phi)}
                            {Y_{\rm Mixed}(\Delta\phi)}
                            \times
\frac{\int Y_{\rm Mixed}(\Delta\phi)d\Delta\phi}
                            {\int Y_{\rm Same}(\Delta\phi)d\Delta\phi}                      
\label{Eq:CF_defined}
\end{equation}
where $Y_{\rm Same}(\Delta\phi=\phi_{2}-\phi_{1})$ and $Y_{\rm Mixed}(\Delta\phi)$ are, respectively, the 
numbers of particle pairs (i.e. one particle is in bin $1$ and the other particle is in bin $2$) at a given $\Delta\phi$ and within a given $p_{\rm{T}}$ range.  This definition of $C(\Delta \phi)$ removes a trivial dependence on the number of produced particles in both bins~\cite{Adler:2005ee,Adare:2014keg}.

In this exercise we calculate for p+Pb events with $N_{\rm track}>110$ (measured in $|\eta |<2.4$ and $p_{\rm{T}}>0.4$ GeV/$c$) and for pairs of charged particles with $1<p_{\rm{T}}<2$ GeV/$c$. To illustrate the effect we choose five different $\Sigma \eta$ configurations for a given $\Delta \eta \sim$ 4: (i) bins 1 and 2 are respectively given by $[-6.2,-5.8]$ and $[-2.2,-1.8]$ for $\Sigma \eta \sim -8$, (ii) $[-4.2, -3.8]$ and $[-0.2,0.2]$ for $\Sigma \eta \sim -4$, (iii) $[-2.2,-1.8]$ and $[1.8, 2.2]$ for $\Sigma \eta \sim 0$, (iv) $[-0.2,0.2]$ and $[3.8, 4.2]$ for $\Sigma \eta \sim 4$, and (v) $[1.8,2.2]$ and $[5.8,6.2]$ for $\Sigma \eta \sim 8$. In our calculations, a Pb nucleus is characterized by a positive $\eta$, which means that the increasing value of $\Sigma \eta$ corresponds to shifting towards a Pb fragmentation region. 

\begin{figure}[t]
\begin{center}
\includegraphics[scale=0.4]{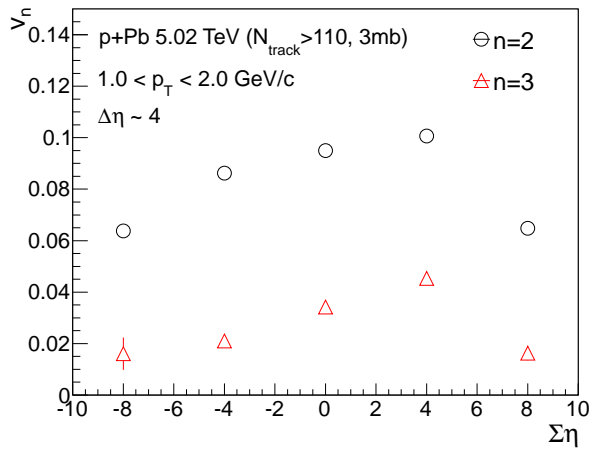}
\end{center}
\par
\vspace{-5mm}
\caption{The second and the third Fourier coefficients, as a function of the pseudorapidity sum $\Sigma \eta = \eta_1 + \eta_2$ at a given pseudorapidity separation $\Delta \eta \sim$ 4. Increasing $\Sigma \eta$ corresponds to shifting towards a Pb-nucleus fragmentation region.}
\label{fig:rap:v2:v3}
\end{figure}

To illustrate the effect we extract the second and the third Fourier coefficients of $C(\Delta\phi)$ 
\begin{equation}
C(\Delta \phi )=1+\sum_{n} 2v_{n}^{2}\cos (n \Delta \phi )
\end{equation}
and plot them as a function of $\Sigma \eta$. This result is presented in Fig. \ref{fig:rap:v2:v3}. Both $v_2$ and $v_3$ increase gradually when going from a proton side to a Pb-nucleus side. This result is expected since on a Pb-going side we have significantly more produced partons and final particles. Another possible reason is the expected difference between the forward and backward eccentricities \cite{Jia:2014ysa}. As expected, far in a nucleus fragmentation region both $v_2$ and $v_3$ start decreasing towards zero.

\begin{figure}
\begin{center}
\includegraphics[scale=0.4]{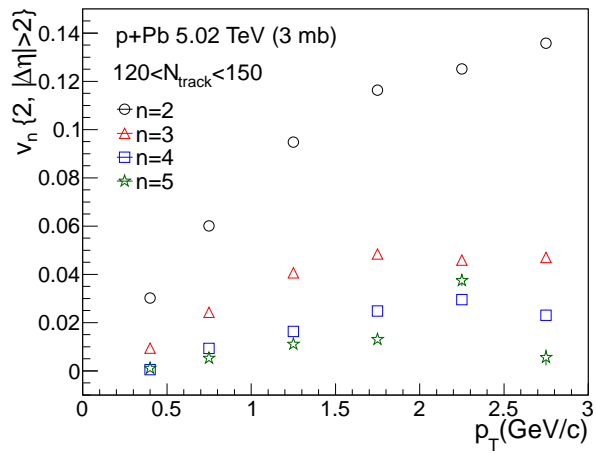}
\end{center}
\par
\vspace{-5mm}
\caption{The AMPT results for the two-particle azimuthal correlation function Fourier coefficients as a function of the transverse momentum in p+Pb collisions.}
\label{fig:v4_v5_pt}
\end{figure}

Finally, in Fig. \ref{fig:v4_v5_pt} we present our predictions for the higher-order Fourier coefficients, $v_4$ and $v_5$, in p+Pb collisions. In the AMPT model with the string melting mechanism, both $v_4$ and $v_5$ are roughly a factor of $2$ smaller than the $v_3$ coefficient. In our plot we only show the results for one centrality class $120<N_{\rm track}<150$ however, similar to $v_2$ and $v_3$ presented in Fig. \ref{fig:v2_v3_pt}, the results for $v_4$ and $v_5$ weakly change with different $N_{\rm track}$ classes.

Before concluding the paper we offer several comments.

Our results suggest that the incoherent scattering of partons plays an important role in the early stage of p+Pb and peripheral Pb+Pb collisions. Moreover, as discussed in Ref. \cite{Ma:2014pva}, the AMPT model allows to understand the ridge effect in p+p for all measured $N_{\rm track}$ and $p_{\rm{T}}$. It is a nontrivial fact that all features present in the data can be qualitatively and quantitatively reproduced within a simple AMPT model.

We checked that the average number of elastic scatterings per parton is approximately $2$ for $N_{\rm track}=200$ in p+Pb, and changes monotonically with $N_{\rm track}$. We find it interesting that such a small number of collisions is sufficient to reproduce the data. 

In our approach we assume that partons scatter incoherently. The lifetime of the partonic matter is roughly 1 fm/$c$ (the time when partons stop interacting) and one could question the validity of this assumption. A simple estimate suggests that it is not unjustified. $\sigma=3$ mb corresponds to the area of 0.3 fm$^2$. $N_{\rm track}$ = 200 corresponds to roughly 40 particles per unit of rapidity and to the effective area per parton of $\sim 0.1$ fm$^2$ (we take the radius of p+Pb to be 2 fm). This number is of the same order of magnitude as $\sigma$ indicating that in a parton's interaction area there are only a few partons (it is consistent with a small number of elastic scatterings), which makes our assumption plausible. The success of our approach could serve as an additional argument in favor of this assumption.

Finally we note that the effect of the hadronic cascade, which can be switched on and off in our approach, has a negligible effect on our results.

In conclusion, using the AMPT model with the string melting mechanizm, we investigated the elliptic and triangular Fourier coefficients of the long-range two-particle azimuthal correlation function in p+Pb and peripheral Pb+Pb collisions. In this model all initial minijets and soft strings are converted into partons which subsequently undergo elastic scatterings. This mechanizm allows us to understand various "flow" data measured in p+Pb and Pb+Pb collisions. In particular we obtained a good description of $v_2(p_{\rm{T}})$ and $v_3(p_{\rm{T}})$ in p+Pb for a broad range of the transverse momentum and for various centrality classes. The dependence of the integrated $v_2$ and $v_3$ on the number of produced charged particles, $N_{\rm track}$, is also nicely reproduced. In peripheral Pb+Pb collisions $v_3(p_{\rm{T}})$ and the integrated $v_3$ coefficients are in satisfactory agreement with the CMS data, however, $v_2$ is underestimated for higher transverse momentum resulting in $20\%$ disagreement for the integrated $v_2$. We also verified the mass ordering of $v_2$ for pions, kaons and protons. We further predicted the pseudorapidity dependence of the two-particle azimuthal correlation function. We observed that $v_2$ and $v_3$ are gradually growing when going from a proton side to a Pb-nucleus side. Finally we calculated the higher-order Fourier coefficients, $v_4$ and $v_5$, in p+Pb collisions and found them to be about a factor of $2$ smaller than the $v_3$ coefficient. 
We hope that the results presented in this Letter will allow to disentangle between competing models of p+A collisions.

\bigskip

Discussions with L. McLerran and Sheng-Li Huang are appreciated. 
A.B. is supported through the RIKEN-BNL Research Center and grant No. UMO-2013/09/B/ST2/00497. 
G.-L. M. is supported by the Major State Basic Research Development Program in China under Grant No. 2014CB845404, the National Natural Science Foundation of China under Grants No. 11175232, No. 11375251, No. 11035009, and No. 11421505, the Knowledge Innovation Program of Chinese Academy of Sciences under Grant No. KJCX2-EW-N01, the Innovation Fund of Key Laboratory of Quark and Lepton Physics (Central China Normal University) under Grant No. QLPL2011P01, and the “Shanghai Pujiang Program” under Grant No. 13PJ1410600.


\begin{thebibliography}{99}

%\cite{Ma:2014pva}
\bibitem{Ma:2014pva} 
  G.~L.~Ma and A.~Bzdak,
  %``Long-range azimuthal correlations in proton–proton and proton–nucleus collisions from the incoherent scattering of partons,''
  Phys.\ Lett.\ B {\bf 739}, 209 (2014).
%  [arXiv:1404.4129 [hep-ph]].
  %%CITATION = ARXIV:1404.4129;%%
  
%\cite{Lin:2004en}
\bibitem{Lin:2004en} 
  Z.~-W.~Lin, C.~M.~Ko, B.~-A.~Li, B.~Zhang and S.~Pal,
  %``A Multi-phase transport model for relativistic heavy ion collisions,''
  Phys.\ Rev.\ C {\bf 72}, 064901 (2005).
%  [nucl-th/0411110].
  %%CITATION = NUCL-TH/0411110;%%
  
%\cite{Khachatryan:2010gv}
\bibitem{Khachatryan:2010gv} 
  V.~Khachatryan {\it et al.}  [CMS Collaboration],
  %``Observation of Long-Range Near-Side Angular Correlations in Proton-Proton Collisions at the LHC,''
  JHEP {\bf 1009}, 091 (2010).
%  [arXiv:1009.4122 [hep-ex]].
  %%CITATION = ARXIV:1009.4122;%%
  
%\cite{CMS:2012qk}
\bibitem{CMS:2012qk}
  S.~Chatrchyan {\it et al.}  [CMS Collaboration],
  %``Observation of long-range near-side angular correlations in proton-lead collisions at the LHC,''
  Phys.\ Lett.\ B {\bf 718} (2013) 795.
%  [arXiv:1210.5482 [nucl-ex]].
  %%CITATION = ARXIV:1210.5482;%%
   
%\cite{Chatrchyan:2013nka}
\bibitem{Chatrchyan:2013nka} 
  S.~Chatrchyan {\it et al.}  [CMS Collaboration],
  %``Multiplicity and transverse momentum dependence of two- and four-particle correlations in pPb and PbPb collisions,''
  Phys.\ Lett.\ B {\bf 724}, 213 (2013).
%  [arXiv:1305.0609 [nucl-ex]].
  %%CITATION = ARXIV:1305.0609;%%
  
%\cite{Abelev:2012ola}
\bibitem{Abelev:2012ola} 
  B.~Abelev {\it et al.}  [ALICE Collaboration],
  %``Long-range angular correlations on the near and away side in $p$-Pb collisions at $\sqrt{s_{NN}}=5.02$ TeV,''
  Phys.\ Lett.\ B {\bf 719}, 29 (2013).
%  [arXiv:1212.2001].
  %%CITATION = ARXIV:1212.2001;%%
  
%\cite{ABELEV:2013wsa}
\bibitem{ABELEV:2013wsa} 
  B.~B.~Abelev {\it et al.}  [ALICE Collaboration],
  %``Long-range angular correlations of pi, K and p in p--Pb collisions at sqrt(s_NN) = 5.02 TeV,''
  Phys.\ Lett.\ B {\bf 726}, 164 (2013).
%  [arXiv:1307.3237 [nucl-ex]].
  %%CITATION = ARXIV:1307.3237;%%
  
%\cite{Aad:2012gla}
\bibitem{Aad:2012gla} 
  G.~Aad {\it et al.}  [ATLAS Collaboration],
  %``Observation of Associated Near-side and Away-side Long-range Correlations in $\sqrt{s_{NN}}$=5.02 TeV Proton-lead Collisions with the ATLAS Detector,''
  Phys.\ Rev.\ Lett.\  {\bf 110}, 182302 (2013).
%  [arXiv:1212.5198 [hep-ex]].
  %%CITATION = ARXIV:1212.5198;%%
  
%\cite{Aad:2013fja}
\bibitem{Aad:2013fja} 
  G.~Aad {\it et al.}  [ATLAS Collaboration],
  %``Measurement with the ATLAS detector of multi-particle azimuthal correlations in p+Pb collisions at sqrt(s_NN)=5.02 TeV,''
  Phys.\ Lett.\ B {\bf 725}, 60 (2013).
%  [arXiv:1303.2084 [hep-ex]].
  %%CITATION = ARXIV:1303.2084;%%
  
%\cite{Adare:2013piz}
\bibitem{Adare:2013piz} 
  A.~Adare {\it et al.}  [PHENIX Collaboration],
  %``Quadrupole anisotropy in dihadron azimuthal correlations in central d+Au collisions at sqrt(s_NN)=200 GeV,''
  Phys.\ Rev.\ Lett.\  {\bf 111}, 212301 (2013).
%  [arXiv:1303.1794 [nucl-ex]].
  %%CITATION = ARXIV:1303.1794;%%

%\cite{Bozek:2011if}
\bibitem{Bozek:2011if} 
  P.~Bozek,
  %``Collective flow in p-Pb and d-Pd collisions at TeV energies,''
  Phys.\ Rev.\ C {\bf 85}, 014911 (2012).
%  [arXiv:1112.0915 [hep-ph]].
  %%CITATION = ARXIV:1112.0915;%%
 
%\cite{Shuryak:2013ke}
\bibitem{Shuryak:2013ke} 
  E.~Shuryak and I.~Zahed,
  %``High Multiplicity pp and pA Collisions: Hydrodynamics at its Edge and Stringy Black Hole,''
  Phys.\ Rev.\ C {\bf 88}, 044915 (2013).
%  [arXiv:1301.4470 [hep-ph]].
  %%CITATION = ARXIV:1301.4470;%%
  
%\cite{Bzdak:2013zma}
\bibitem{Bzdak:2013zma} 
  A.~Bzdak, B.~Schenke, P.~Tribedy and R.~Venugopalan,
  %``Initial state geometry and the role of hydrodynamics in proton-proton, proton-nucleus and deuteron-nucleus collisions,''
  Phys.\ Rev.\ C {\bf 87}, no. 6, 064906 (2013).
%  [arXiv:1304.3403 [nucl-th]].
  %%CITATION = ARXIV:1304.3403;%%
  
%\cite{Bozek:2013uha}
\bibitem{Bozek:2013uha} 
  P.~Bozek and W.~Broniowski,
  %``Collective dynamics of the high-energy proton-nucleus collisions,''
  Phys.\ Rev.\ C {\bf 88}, 014903 (2013).
%  [arXiv:1304.3044 [nucl-th]].
  %%CITATION = ARXIV:1304.3044;%%
  
%\cite{Qin:2013bha}
\bibitem{Qin:2013bha} 
  G.~-Y.~Qin and B.~Mueller,
  %``Elliptic and triangular flow anisotropy in deuteron-gold collisions at RHIC and proton-lead collisions at the LHC,''
  Phys.\ Rev.\ C {\bf 89}, 044902 (2014).
%  [arXiv:1306.3439 [nucl-th]].
  %%CITATION = ARXIV:1306.3439;%%
  
%\cite{Werner:2013tya}
\bibitem{Werner:2013tya} 
  K.~Werner, B.~Guiot, I.~Karpenko and T.~Pierog,
  %``Analysing radial flow features in p-Pb and p-p collisions at several TeV by studying identified particle production in EPOS3,''
  Phys.\ Rev.\ C {\bf 89}, 064903 (2014).
%  [arXiv:1312.1233 [nucl-th]].
  %%CITATION = ARXIV:1312.1233;%%
  
%\cite{Kozlov:2014fqa}
\bibitem{Kozlov:2014fqa} 
  I.~Kozlov, M.~Luzum, G.~Denicol, S.~Jeon and C.~Gale,
  %``Transverse momentum structure of pair correlations as a signature of collective behavior in small collision systems,''
  arXiv:1405.3976 [nucl-th].
  %%CITATION = ARXIV:1405.3976;%%
  
%\cite{Bzdak:2013rya}
\bibitem{Bzdak:2013rya} 
  A.~Bzdak, P.~Bozek and L.~McLerran,
  %``Fluctuation induced symmetry breaking and the equality of multi-particle eccentricities for four or more particles,''
  Nucl.\ Phys.\ A {\bf 927}, 15 (2014).
%  [arXiv:1311.7325 [hep-ph]].
  %%CITATION = ARXIV:1311.7325;%%
  
%\cite{Yan:2013laa}
\bibitem{Yan:2013laa} 
  L.~Yan and J.~-Y.~Ollitrault,
  %``Universal fluctuation-driven eccentricities in proton-proton, proton-nucleus and nucleus-nucleus collisions,''
  Phys.\ Rev.\ Lett.\  {\bf 112}, 082301 (2014).
%  [arXiv:1312.6555 [nucl-th]].
  %%CITATION = ARXIV:1312.6555;%%
  
%\cite{Bzdak:2013raa}
\bibitem{Bzdak:2013raa} 
  A.~Bzdak and V.~Skokov,
  %``Multi-particle eccentricities in collisions dominated by fluctuations,''
  arXiv:1312.7349 [hep-ph].
  %%CITATION = ARXIV:1312.7349;%%
  
%\cite{CMS:v4-v8:qm14:talk}
\bibitem{CMS:v4-v8:qm14:talk} 
  CMS Collaboration, CMS Physics Analysis Summary No. CMS-PAS-HIN-14-006, 2014, http://cds.cern.ch/record/1705485

%\cite{Bozek:2013ska}
\bibitem{Bozek:2013ska} 
  P.~Bozek, W.~Broniowski and G.~Torrieri,
  %``Mass hierarchy in identified particle distributions in proton-lead collisions,''
  Phys.\ Rev.\ Lett.\  {\bf 111}, 172303 (2013).
%  [arXiv:1307.5060 [nucl-th]].
  %%CITATION = ARXIV:1307.5060;%%
  
%\cite{Werner:2013ipa}
\bibitem{Werner:2013ipa} 
  K.~Werner, M.~Bleicher, B.~Guiot, I.~Karpenko and T.~Pierog,
  %``Evidence for flow in pPb collisions at 5 TeV from v2 mass splitting,''
  Phys.\ Rev.\ Lett.\  {\bf 112}, 232301 (2014).
%  [arXiv:1307.4379 [nucl-th]].
  %%CITATION = ARXIV:1307.4379;%%

%\cite{Gelis:2010nm}
\bibitem{Gelis:2010nm} 
  F.~Gelis, E.~Iancu, J.~Jalilian-Marian and R.~Venugopalan,
  %``The Color Glass Condensate,''
  Ann.\ Rev.\ Nucl.\ Part.\ Sci.\  {\bf 60}, 463 (2010).
%  [arXiv:1002.0333 [hep-ph]].
  %%CITATION = ARXIV:1002.0333;%%

% both pp and pPb 
%\cite{Dusling:2013oia}
\bibitem{Dusling:2013oia} 
  K.~Dusling and R.~Venugopalan,
  %``Comparison of the Color Glass Condensate to di-hadron correlations in proton-proton and proton-nucleus collisions,''
  Phys.\ Rev.\ D {\bf 87}, 094034 (2013).
%  [arXiv:1302.7018 [hep-ph]].
  %%CITATION = ARXIV:1302.7018;%%

%\cite{Kovchegov:2012nd}
\bibitem{Kovchegov:2012nd} 
  Y.~V.~Kovchegov and D.~E.~Wertepny,
  %``Long-Range Rapidity Correlations in Heavy-Light Ion Collisions,''
  Nucl.\ Phys.\ A {\bf 906}, 50 (2013).
%  [arXiv:1212.1195].
  %%CITATION = ARXIV:1212.1195;%%

%\cite{Kovner:2012jm}
\bibitem{Kovner:2012jm} 
  A.~Kovner and M.~Lublinsky,
  %``Angular and long range rapidity correlations in particle production at high energy,''
  Int.\ J.\ Mod.\ Phys.\ E {\bf 22}, 1330001 (2013).
%  [arXiv:1211.1928 [hep-ph]].
  %%CITATION = ARXIV:1211.1928;%%
  
%\cite{Bzdak:2013lva}
\bibitem{Bzdak:2013lva} 
  A.~Bzdak and V.~Skokov,
  %``Average transverse momentum of hadrons in proton-nucleus collisions in the wounded nucleon model,''
  Phys.\ Lett.\ B {\bf 726}, 408 (2013).
%  [arXiv:1306.5442 [nucl-th]].
  %%CITATION = ARXIV:1306.5442;%%
  
%\cite{Bzdak:2013zla}
\bibitem{Bzdak:2013zla} 
  A.~Bzdak and V.~Skokov,
  %``Decisive test of color coherence in proton-nucleus collisions at the LHC,''
  Phys.\ Rev.\ Lett.\  {\bf 111}, 182301 (2013).
%  [arXiv:1307.6168 [hep-ph]].
  %%CITATION = ARXIV:1307.6168;%%
  
%\cite{Bozek:2013sda}
\bibitem{Bozek:2013sda} 
  P.~Bozek, A.~Bzdak and V.~Skokov,
  %``The rapidity dependence of the average transverse momentum in p+Pb collisions at the LHC: the Color Glass Condensate versus hydrodynamics,''
  Phys.\ Lett.\ B {\bf 728}, 662 (2014).
%  [arXiv:1309.7358 [hep-ph]].
  %%CITATION = ARXIV:1309.7358;%%
  
%\cite{Basar:2013hea}
\bibitem{Basar:2013hea} 
  G.~Başar and D.~Teaney,
  %``Scaling relation between pA and AA collisions,''
  Phys.\ Rev.\ C {\bf 90}, 054903 (2014).
%  [arXiv:1312.6770 [nucl-th]].
  %%CITATION = ARXIV:1312.6770;%%

%\cite{Konchakovski:2014wqa}
\bibitem{Konchakovski:2014wqa} 
  V.~P.~Konchakovski, W.~Cassing and V.~D.~Toneev,
  %``p-Pb collisions at 5.02 TeV in the Parton-Hadron-String-Dynamics transport approach,''
  J.\ Phys.\ G {\bf 41}, 105004 (2014).
%  [arXiv:1401.4409 [nucl-th]].
  %%CITATION = ARXIV:1401.4409;%%
  %3 citations counted in INSPIRE as of 16 Sep 2014

%\cite{Sickles:2013yna}
\bibitem{Sickles:2013yna} 
  A.~M.~Sickles,
  %``Possible Evidence for Radial Flow of Heavy Mesons in d+Au Collisions,''
  Phys.\ Lett.\ B {\bf 731}, 51 (2014).
%  [arXiv:1309.6924 [nucl-th]].
  %%CITATION = ARXIV:1309.6924;%%
  
%\cite{Noronha:2014vva}
\bibitem{Noronha:2014vva} 
  J.~Noronha and A.~Dumitru,
  %``Azimuthal asymmetries in high-energy collisions of protons with holographic shockwaves,''
  Phys.\ Rev.\ D {\bf 89}, 094008 (2014).
%  [arXiv:1401.4467 [hep-ph]].
  %%CITATION = ARXIV:1401.4467;%%
  
%\cite{Floerchinger:2014fta}
\bibitem{Floerchinger:2014fta} 
  S.~Floerchinger and U.~A.~Wiedemann,
  %``Statistics of initial density perturbations in heavy ion collisions and their fluid dynamic response,''
  JHEP {\bf 1408}, 005 (2014).
%  [arXiv:1405.4393 [hep-ph]].
  %%CITATION = ARXIV:1405.4393;%%
  
% triangularity
%\cite{Alver:2010gr}
\bibitem{Alver:2010gr} 
  B.~Alver and G.~Roland,
  %``Collision geometry fluctuations and triangular flow in heavy-ion collisions,''
  Phys.\ Rev.\ C {\bf 81}, 054905 (2010)
  [Erratum-ibid.\ C {\bf 82}, 039903 (2010)].
%  [arXiv:1003.0194 [nucl-th]].
  %%CITATION = ARXIV:1003.0194;%%
   
%\cite{Ma:2014xfa}
\bibitem{Ma:2014xfa} 
  L.~Ma, G.~L.~Ma and Y.~G.~Ma,
  %``Anisotropic flow and flow fluctuations for Au + Au at $\sqrt{s_{NN}}$ = 200 GeV in a multiphase transport model,''
  Phys.\ Rev.\ C {\bf 89}, 044907 (2014).
%  [arXiv:1404.5935 [nucl-th]].
  %%CITATION = ARXIV:1404.5935;%%

%\cite{Xu:2011jm}
\bibitem{Xu:2011jm} 
  J.~Xu and C.~M.~Ko,
  %``Higher-order anisotropic flows and dihadron correlations in Pb-Pb collisions at $\sqrt{s_{NN}}=2.76$ TeV in a multiphase transport model,''
  Phys.\ Rev.\ C {\bf 84}, 044907 (2011).
%  [arXiv:1108.0717 [nucl-th]].
  %%CITATION = ARXIV:1108.0717;%%

\bibitem{Ma:2010dv} 
  G.~-L.~Ma and X.~-N.~Wang,
  %``Jets, Mach cone, hot spots, ridges, harmonic flow, dihadron and $\gamma$-hadron correlation in high-energy heavy-ion collisions,''
  Phys.\ Rev.\ Lett.\  {\bf 106}, 162301 (2011).
%  [arXiv:1011.5249 [nucl-th]].
  %%CITATION = ARXIV:1011.5249;%%

\bibitem{footnote}  
  We checked that varying the cross-section in Pb+Pb only slightly improves the situation. At present we do not have an explanation why $v_2$ in Pb+Pb is underestimated. We can only speculate that it is related to the large average ellipticity present in peripheral Pb+Pb which could be underestimated in the Glauber model. Another possibility is some additional contribution to $v_2$, e.g., because of to an interplay of jet quenching and a strong elliptical shape of peripheral Pb+Pb.

\bibitem{Adler:2005ee} 
  S.~S.~Adler {\it et al.}  [PHENIX Collaboration],
  %``Dense-Medium Modifications to Jet-Induced Hadron Pair Distributions in Au+Au Collisions at s(NN)**(1/2) = 200-GeV,''
  Phys.\ Rev.\ Lett.\  {\bf 97}, 052301 (2006).
%  [nucl-ex/0507004].
  %%CITATION = NUCL-EX/0507004;%%
  
\bibitem{Adare:2014keg} 
  A.~Adare {\it et al.}  [PHENIX Collaboration],
  %``Measurement of long-range angular correlation and quadrupole anisotropy of pions and (anti)protons in central $d$$+$Au collisions at $\sqrt{s_{_{NN}}}$=200 GeV,''
  arXiv:1404.7461 [nucl-ex].
  %%CITATION = ARXIV:1404.7461;%%
  
%\cite{Jia:2014ysa}
\bibitem{Jia:2014ysa} 
  J.~Jia and P.~Huo,
  %``Forward-backward eccentricity and participant-plane angle fluctuations and their influences on longitudinal dynamics of collective flow,''
  Phys.\ Rev.\ C {\bf 90}, 034915 (2014).
%  [arXiv:1403.6077 [nucl-th]].
  %%CITATION = ARXIV:1403.6077;%%
  
\end{thebibliography}
\end{document}